\documentclass[aps,pre,twocolumn,groupedaddress,floatfix,10pt]{revtex4-2}

\usepackage{amsmath}
\usepackage{graphicx}
\usepackage{booktabs}
\usepackage[usenames,dvipsnames]{color}
\usepackage{tabularx}

\begin{document}


\title{Kinetic temperatures and inertial effects in a nonequilibrium bead-spring model}


\author{Jetin E Thomas}
\email[]{jetinthomas@iisermohali.ac.in}
\affiliation{Department of Physical Sciences, Indian Institute of Science Education and Research Mohali, \\ Knowledge city, Sector 81, Manauli, PO, Sahibzada Ajit Singh Nagar, Punjab 140306, India}
\author{Ramandeep S. Johal}
\email[]{rsjohal@iisermohali.ac.in}
\affiliation{Department of Physical Sciences, Indian Institute of Science Education and Research Mohali, \\ Knowledge city, Sector 81, Manauli, PO, Sahibzada Ajit Singh Nagar, Punjab 140306, India}


\date{\today}

\begin{abstract}
We investigate a nonequilibrium steady-state model consisting of two coupled beads with arbitrary masses 
in contact with two thermal baths at different temperatures. 
Using a covariance-matrix approach together with numerical simulations of the underdamped Langevin dynamics, we characterize steady-state probability distributions, heat transport, and entropy production. 
We show that irreversibility measures such as entropy production and heat current 
are invariant under an exchange of the bead masses, whereas energy-storage observables depend explicitly on the mass arrangement in a symmetrical set up. 
This reveals a fundamental distinction: energy observables exhibit path dependence in singular mass limits, while transport and irreversibility remain well defined.  
We show that kinetic temperatures provide the natural variables governing the thermodynamics of the system: their difference controls transport and entropy production, while their sum determines the mean energy via a model specific generalized equipartition relation. 
In the infinite-mass limit, only constitutive relations expressed in terms of kinetic temperatures remain meaningful. 
Thus, energy, transport, and irreversibility are unified through kinetic temperatures as the organizing variables. 
We also derive an effective temperature that defines an equilibrium-like canonical distribution. 
Finally, we analyze the notion of ergodicity and show that the time-averaged observables converge significantly faster than the ensemble averages.
\end{abstract}


\maketitle

\section{\label{Introduction} Introduction}

Bead--spring models coupled to multiple heat baths serve as minimal and analytically tractable models of nonequilibrium steady states (NESS).
Originally developed in polymer physics, most notably in the Rouse model \cite{rouse1953theory}, 
they were later extended to nonequilibrium settings by coupling different degrees of freedom to thermal reservoirs at unequal temperatures.
Since the late 1990s, such models have served as workhorse systems for studying heat conduction, entropy production, and fluctuations in small-scale systems, as they allow exact calculations while exhibiting irreversibility, probability currents, and sustained heat flows \cite{lebowitz1999gallavotti,dhar2008heat}.
Variants with a few particles coupled to multiple heat baths were studied by Van den Broeck, Kawai, Esposito, and collaborators to analyze entropy production, heat currents, and steady-state covariance structures using Lyapunov-based approaches \cite{van2004microscopic}.
More recently, two-bead models have been used to describe phonon heat transfer across a vacuum mediated by quantum fluctuations \cite{fong2019phonon}.

A central difficulty in characterizing NESS lies in extending the notion of temperature beyond equilibrium.
In equilibrium statistical mechanics, temperature is uniquely defined through the Gibbs--Boltzmann distribution, and the equipartition theorem that assigns the same temperature to all quadratic degrees of freedom.
In contrast, nonequilibrium steady states are not generally described by a Boltzmann measure, and different degrees of freedom need not equilibrate, rendering a unique temperature ill-defined \cite{cugliandolo2011effective,seifert2012stochastic}.
This has motivated the introduction of several temperature-like quantities in nonequilibrium contexts, including effective temperatures defined via fluctuation–dissipation relations \cite{Cugliandolo1997,Cugliandolo2011,Seifert2012}, configurational temperatures \cite{Rugh1997,Jepps2000}, and kinetic temperatures based on velocity fluctuations \cite{goldhirsch2003rapid,Kadanoff1999}.
Among these, the kinetic temperature provides a direct measure of energy stored in inertial degrees of freedom, even far from equilibrium \cite{goldhirsch2003rapid}.

For linear Langevin systems, NESS can be fully characterized through their probability distributions.
These distributions can be obtained by solving the associated Fokker--Planck equation or, equivalently, by determining the covariance matrix through Lyapunov equations \cite{herzel1991risken,gardiner1985handbook}.
The covariance matrix therefore determines steady-state fluctuations and correlations \cite{van2004microscopic,seifert2012stochastic,barato2015thermodynamic}.
Recent work has formulated nonequilibrium equations of state for harmonic systems by introducing equilibrium-like thermodynamic quantities \cite{wu2022nonequilibrium}.
In a related approach, Tu \cite{tu2025weighted} defined an effective temperature from steady-state position distributions, reproducing spatial fluctuations without explicit use of covariance matrices.

Nonequilibrium systems also exhibit a separation between different classes of observables.
Early theories emphasized transport coefficients and entropy production as measures of irreversibility \cite{onsager1931reciprocal,deGrootMazur}.
Stochastic thermodynamics further distinguishes entropy production and currents (time-antisymmetric) from energies and static distributions (time-symmetric) \cite{seifert2012stochastic,spinney2012entropy}.
This reflects a decomposition of dynamics into reversible and irreversible components \cite{maes2007entropy}.
Here, we demonstrate this separation explicitly: exchanging unequal bead masses leaves entropy production and transport invariant, while energy-storage observables depend on the mass arrangement.
This provides a minimal setting in which inertia selectively affects the time-symmetric observables without altering irreversibility. We emphasize that this separation is established here within a linear Langevin framework with Gaussian steady states; its extension to nonlinear or Hamiltonian systems is not guaranteed and may depend on additional dynamical constraints.

Different observables approach the overdamped limit differently as the inertia is reduced.
Position distributions may converge smoothly in the zero-mass limit, whereas entropy production can retain dependence on phase-space dynamics \cite{spinney2012entropy,celani2012anomalous}.
Transport properties, governed by correlations and mode structure, can exhibit a behavior distinct from the entropy production \cite{derrida2007nonequilibrium}.
These distinctions become particularly transparent when asymmetries are introduced and kinetic temperatures are used to characterize transport.
From a computational perspective, the characterization of NESS also depends on averaging protocols: ensemble averaging suppresses fluctuations efficiently, while time averaging along long trajectories can be more effective in ergodic systems \cite{khinchin1949mathematical,vanKampen2007}. While these features arise naturally here, they may change in the presence of nonlinear interactions where the steady-state distributions become non-Gaussian and correlations are no longer fully captured by the covariance structure.

Despite extensive studies of two-bath harmonic systems, the distinct roles of inertia in energy storage and irreversibility remain unclear. 
Here, we analyze the two-beads system and compare energy storage, transport, and irreversibility within a unified framework. 
The remainder of the paper is organized as follows. Sec.~\ref{Model} introduces the model. 
Sec.~\ref{Results} presents the steady-state distributions, entropy production, thermal transport, energy, and ergodicity. 
Sec.~\ref{Discussions} provides a unified discussion of these results and their physical implications.

\section{\label{Model} Model}

\begin{figure}
\includegraphics[scale=0.4]{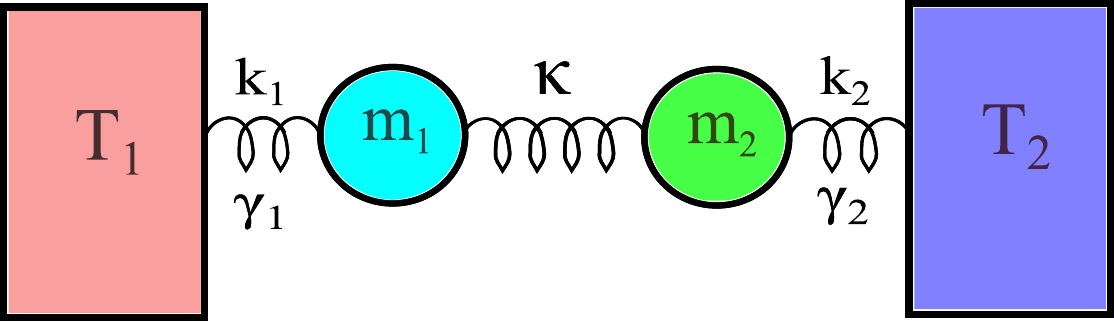}
\caption{Two beads, of mass $m_{1}$ and $m_{2}$, simultaneously 
coupled to each other with a spring having spring constant  $\kappa$ as well as with a bath at temperature $T_{1}$ ($T_{2}$) with spring constant $k_{1}$ ($k_{2}$). The frictional drag coefficient is  $\gamma_{1}$ ($\gamma_{2}$) for the left (right) particle.}
\label{twobeadssetup}
\end{figure}

Our model system consists of two beads, each directly coupled to a heat bath at a different temperature ($T_1 > T_2$) (see Fig. \ref{twobeadssetup}). 
The coupling to the left (right) bath is 
characterized by the spring constant $k_{1} (k_{2})$. Similarly, the harmonic coupling between the beads 
is governed by the spring constant $\kappa$. 

 The Langevin equations \cite{saito2007fluctuation, kundu2011large, fogedby2012heat} for this system are 
\begin{align}
\frac{dx_{1}}{dt} = & \;  v_{1}, \label{pos1_Langevin} \\
\frac{dv_{1}}{dt} = & -\frac{k_{1}x_{1}}{m_{1}}
+\frac{\kappa(x_{2}-x_{1})}{m_{1}} 
-\frac{\gamma_{1}v_{1}}{m_{1}}+\xi_{1}(t), \label{vel1_Langevin}\\
\frac{dx_{2}}{dt} = & \;  v_{2}, \label{pos2_Langevin} \\
\frac{dv_{2}}{dt} = & 
-\frac{k_{2}x_{2}}{m_{2}}
-\frac{\kappa(x_{2}-x_{1})}{m_{2}} 
-\frac{\gamma_{2}v_{2}}{m_{2}}+\xi_{2}(t),
 \label{vel2_Langevin}
\end{align}
where $x_{1}$ ($x_{2}$) and $v_{1}$ ($v_{2}$) are
respectively the displacement from the equilibrium position
and the velocity 
of the bead with mass $m_1$ ($m_2$). The white noise due to a thermal bath satisfies: $\langle \xi_{\alpha}(t) \rangle = 0$ and $\langle \xi_{\alpha}(t)\xi_{\alpha}(t') \rangle = 2{\gamma_{\alpha}T_{\alpha}}\delta(t-t')/m_{\alpha}^2$, with $\alpha=1,2$.
 Here, the physical units for the degrees of freedom and the system parameters can be appropriately chosen to fit the description of the Langevin model, eqs. (\ref{pos1_Langevin}-\ref{vel2_Langevin}).

\section{\label{Results} Results}
\subsection{\label{CPDA_TwoBead}
Steady-state distributions: marginals and the overdamped limit}


Eqs.  
(\ref{pos1_Langevin}-\ref{vel2_Langevin}) represent an Ornstein-Uhlenbeck process where  the probability distribution for positions and velocities of left bead ($x_{1}, v_{1}$) and the right bead ($x_{2},v_{2}$)  is given with the help of the inverse of the covariance matrix ($\boldsymbol{\sigma}^{-1}$), as follows \cite{van2004microscopic, seifert2012stochastic, barato2015thermodynamic}.
\begin{equation}
P(\boldsymbol{z} = [x_{1},x_{2},v_{1},v_{2}]^{T}) = 
\frac{\exp({-{\boldsymbol{z}^{T}\boldsymbol{\sigma}^{-1}\boldsymbol{z}}/{2}})}{\sqrt{{\rm Det}[2\pi \boldsymbol{\sigma}]}}.
\label{Prob_interms_cov_matrix_2B}
\end{equation}
In Appendix A, we have laid out the formulation for the effective temperature ($T_{e}$) from the Langevin equations, which is used to write the joint 
steady-state probability distribution in the form of a Boltzmann-like distribution, as  
\begin{equation}
P(\boldsymbol{z}) = \exp({-(H+\Delta H)/T_{e}})/\mathcal{Z}, 
\end{equation}
where $H$ and $\Delta H$ are given in eqs. (\ref{Hamiltonian_2B}) and (\ref{dH_2B}), respectively. Note that, in general, $\Delta H$ may not be zero and the distributions deviate from the Boltzmann form. The parameter $T_e$ defined through the steady-state distribution is, in general, distinct from the effective temperatures defined via fluctuation–dissipation relations, such as $\Theta = D/\mu_d$ introduced by Hayashi and Takano \cite{hayashi2007temperature}. The latter is based on dynamical response properties, whereas $T_e$ arises from the static probability distribution. In nonequilibrium steady states with multiple reservoirs, fluctuation–dissipation relations are violated, and these different notions of temperature do not coincide. This highlights the absence of a unique defintion of 
temperature in such systems.

\begin{figure}
\includegraphics[scale=0.65]{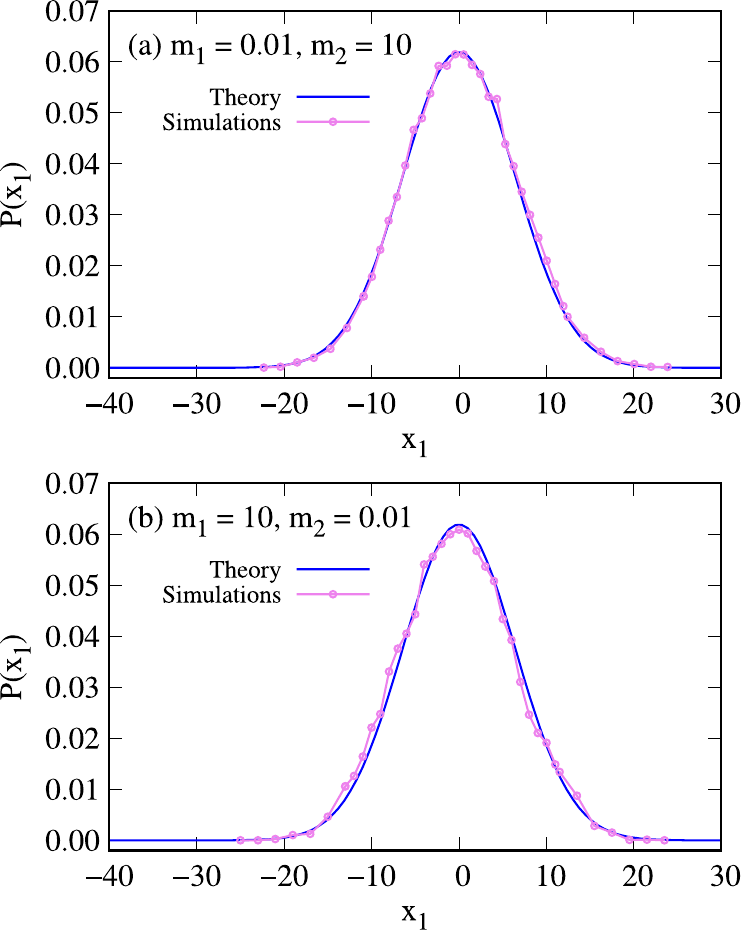}
\caption{The numerically computed marginal probability distribution ($P(x_{1})$) in comparison to the distribution obtained from theory (Eq. (\ref{Prob_interms_cov_matrix_2B})). The distributions are very close to each other for the two cases. The simulations assume $m_{1}=0.01$, $m_{2}=10$ (a), and $m_{1}=10$, $m_{2}=0.01$ (b), $T_{1}=99$, $T_{2}=1$, $\gamma_{1}=\gamma_{2}=1$, $k_{1}=k_{2}=1$, and $\kappa = 2$.}
\label{mar_prob_diffapproach}
\end{figure}

In Fig. \ref{mar_prob_diffapproach}, we include a comparison plot for the marginal  distribution ($P(x_{1})$) in the position space of the first bead ($x_{1}$), derived as $P(x_{1})=\int_{-\infty}^{\infty}dx_{2}\int_{-\infty}^{\infty}dv_{1}\int_{-\infty}^{\infty} dv_{2}P(\boldsymbol{z})$. The distributions from eq. (\ref{Prob_interms_cov_matrix_2B})) overlap with the ones obtained numerically. Also, we do not see any difference in the probability distributions when the asymmetry in the two masses is switched in Fig. \ref{mar_prob_diffapproach}(b) in comparison to Fig. \ref{mar_prob_diffapproach}(a) implying 
that, at steady state, the order in which the masses are placed in the spring-bead network does not matter,
provided the set up is symmetrical in other 
parameters ($k_{1}=k_{2}$ and $\gamma_{1}=\gamma_{2}$).


The marginal distributions in the positions of both beads remain invariant to an exchange between the unequal masses in a symmetrical setup. This invariance relies on the symmetry of the setup and linearity of the dynamics, and need not persist in systems with nonlinear interactions or asymmetric couplings. This could be proved by inferring that in an Ornstein-Uhlenbeck process, we get a Gaussian distribution with zero mean in the steady state \cite{uhlenbeck1930theory}. The variance comes from the relations in eqs. (\ref{Avgx1sq}) and (\ref{Avgx2sq}) which remains unchanged on swapping the unequal masses. On the other hand, the marginal distributions of bead velocities do depend on the order of placing the masses in a symmetrical setup, as seen in eqs. (\ref{Avgv1sq}) and (\ref{Avgv2sq}). These features get encrypted in the mode temperatures ($\propto \langle x_{\alpha}^{2}\rangle$) and kinetic temperatures ($\propto \langle v_{\alpha}^{2}\rangle$), where $\alpha = 1,2$, respectively.

\begin{figure}
\includegraphics[scale=0.65]{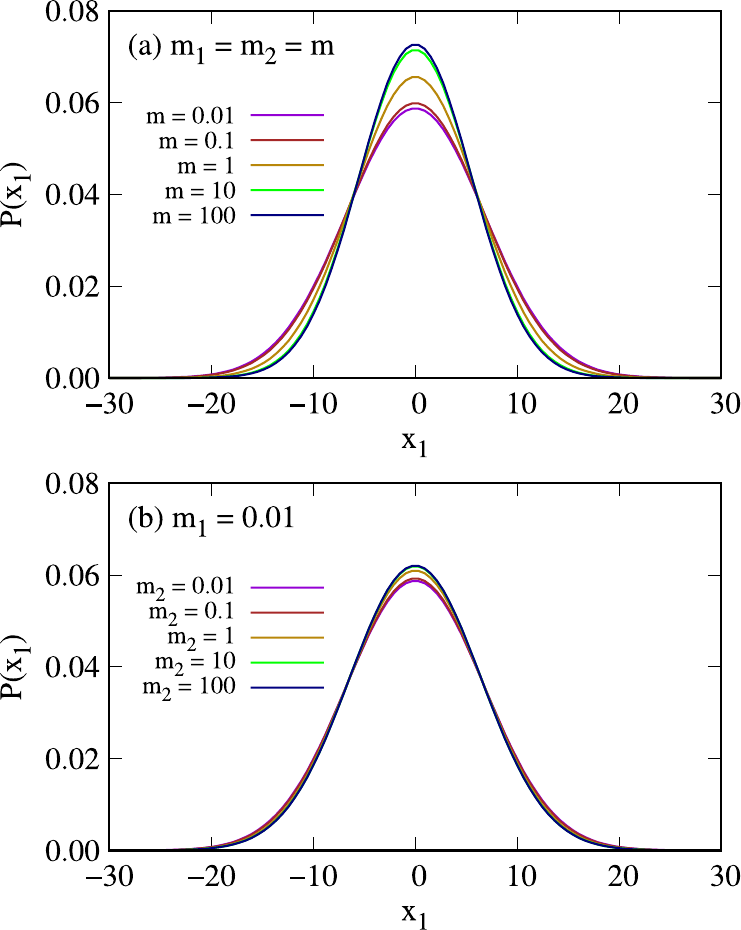}
\caption{The analytical marginal probability distribution ($P(x_{1})$) for different values of the mass $m$ of each bead. a) The distributions become narrower and higher with an increase in mass which is equal for each bead. b) The distributions also increases with increase in mass ($m_{2}$) of the second bead keeping the first bead at fixed mass ($m_{1}=0.01$). But the extent of increase is very small. Here, $T_{1}=99$, $T_{2}=1$, $\gamma_{1}=\gamma_{2}=1$, $k_{1}=k_{2}=1$, and $\kappa = 2$.}
\label{mar_prob_diffmass}
\end{figure}

\begin{figure}
\includegraphics[scale=0.65]{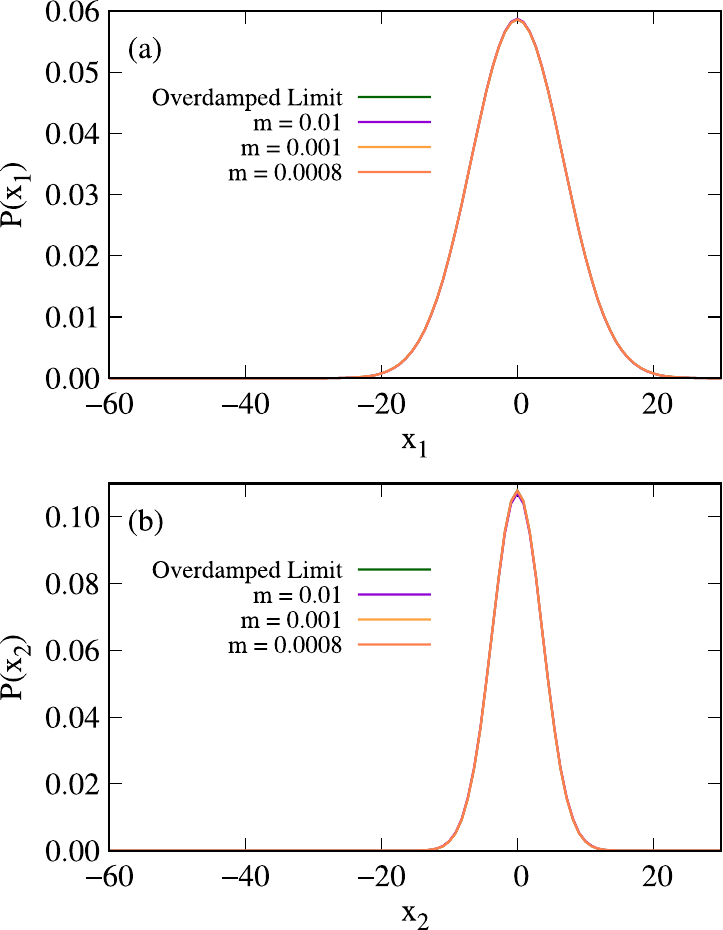}
\caption{The marginal probability distribution for the position ($x_{1}$) of the bead to the left (a) and position ($x_{2}$) of the bead to the right (b). The marginal distributions for the overdamped limit are compared with the analytically obtained distributions of different small masses. The distributions from the overdamped limit and the three small masses lie on top of each other. The curves were drawn at  $T_{1}=99$, $T_{2}=1$, $\gamma_{1}=\gamma_{2}=1$, $k_{1}=k_{2}=1$, and $\kappa = 2$.}
\label{Px1x2_od}
\end{figure}

Figure \ref{mar_prob_diffmass}(a) shows the effect of the mass
parameter where the distribution becomes narrower with an increase in mass, taken equal for both the beads. However, the effect is small if we increase just one mass ($m_{2}$) while keeping the other fixed, as in Fig. \ref{mar_prob_diffmass}(b) on a symmetrical setup. This indicates that increasing the inertia while preserving the symmetry decreases the fluctuations more effectively in comparison to increasing the asymmetry due to an increase in mass of one bead only. Thus, mass asymmetry primarily redistributes fluctuations between degrees of freedom without strongly reshaping the covariance in the position of the bead whose mass is kept constant.
When mass tends to zero, the steady-state overdamped probability distribution ($P_{\rm od}(x_{1},x_{2})$) is of the form
\begin{equation}
\resizebox{0.9\columnwidth}{!}{$P_{\rm od}(x_{1},x_{2}) = \frac{1}{2\pi \sqrt{{\rm Det}[\boldsymbol{\sigma}_{\rm od}]}}\exp\left({-\frac{[x_{1},x_{2}]\boldsymbol{\sigma}_{\rm od}^{-1}[x_{1},x_{2}]^{T}}{2}}\right)$,}
\label{Prob_OD_TwoBead}
\end{equation}
where $\boldsymbol{\sigma}_{\rm od}$ satistfies the Lyapunov equation: $\boldsymbol{F.\sigma}_{\rm od}+\boldsymbol{\sigma}_{\rm od}.\boldsymbol{F}^{T} = 2\boldsymbol{D}$ \cite{maes2003time, seifert2008stochastic, parrondo2009entropy, tome2015stochastic}. The form of $\boldsymbol{F}$ could be derived from the Langevin equations (\ref{vel1_Langevin}) and (\ref{vel2_Langevin}) after 
taking the $m_{\alpha} \rightarrow 0$ limit. This Langevin equation, at the overdamped limit, could be compactly written as
\begin{equation}
\frac{d[x_{1},x_{2}]^{T}}{dt} = - \boldsymbol{F}.[x_{1},x_{2}]^{T} + [\xi_{1}(t)/\gamma_{1},\xi_{2}(t)/\gamma_{2}]^{T}, \label{vec_od_Langevin}
\end{equation}
where $\boldsymbol{F} =  \begin{bmatrix}
\frac{k_{1}+\kappa}{\gamma_{1}} & -\frac{\kappa}{\gamma_{1}}  \\
-\frac{\kappa}{\gamma_{2}} & \frac{k_{2}+\kappa}{\gamma_{2}}
\end{bmatrix}$ and $\boldsymbol{D} =  \begin{bmatrix}
\frac{T_{1}}{\gamma_{1}} & 0  \\
0 & \frac{T_{2}}{\gamma_{2}}
\end{bmatrix}$. 
By assuming $\boldsymbol{\sigma}_{\rm od} =  \begin{bmatrix}
a & b  \\
b & c
\end{bmatrix}$, 
we solve the Lyapunov equation to obtain
\begin{align}
a &= \frac{T_{1}+b\kappa}{k_{1}+\kappa}, \quad 
c = \frac{T_{2}+b\kappa}{k_{2}+\kappa},\\
b & =\frac{\kappa \left(\frac{T_{1}}{\gamma_{2}(k_{1}+\kappa)}+\frac{T_{2}}{\gamma_{1}(k_{2}+\kappa)} \right)}{\left( (k_{1}+\kappa)(k_{2}+\kappa)-\kappa^{2} \right)\left(\frac{1}{\gamma_{2}(k_{1}+\kappa)}+\frac{1}{\gamma_{1}(k_{2}+\kappa)} \right)}.
\end{align}
We have compared the marginal distributions for positions of both the beads, given by $P(x_{\alpha'}) = \int dx_{\alpha}P_{\rm od}(x_{1},x_{2})$, 
$\alpha' \neq \alpha$, with the analytically obtained marginal distributions at very small masses, as in Fig. \ref{Px1x2_od}. We see that the distributions are very close  to each other for all the cases, thus satisfying the definition of probability distribution in the  overdamped limit.

\subsection{\label{Ent_TwoBead} Entropy Production}
The heat fluxes exiting the hot bath and entering
the cold bath are respectively given by \cite{sekimoto2010stochastic, seifert2012stochastic,spinney2012entropy}
\begin{gather}
\dot{Q}_{1} = \gamma_{1} \left( {T_{1}}/{m_{1}}-\langle v_{1}^{2} \rangle \right) = \gamma_{1}(T_{1}-T_{v_{1}})/m_{1},\label{heatrate_left_2Bead}\\
\dot{Q}_{2} = \gamma_{2}\left( {T_{2}}/{m_{2}}-\langle v_{2}^{2} \rangle \right) = \gamma_{2}(T_{2}-T_{v_{2}})/m_{2},\label{heatrate_right_2Bead}
\end{gather}
where $T_{v_{\alpha}}=m_{\alpha}\langle v_{\alpha}^{2}\rangle$  is defined as the kinetic temperature for bead $\alpha=1,2$. Note that 
the kinetic temperature of a bead in NESS is defined 
locally from its mean kinetic energy and is not, in general, analogous to the thermodynamic temperature of the equilibrated system. It becomes equal to the thermodynamic temperature or the bath temperature only when the temperatures of the heat baths are equal and the system reaches equilibrium, as seen using eqs. \ref{Avgv1sq} and \ref{Avgv2sq}. We see that the heat flux between the heat bath and the bead is directly proportional to the difference of the bath temperature and the bead kinetic temperature. From the covariance matrix relations eqs. (\ref{Avgv1sq}) and (\ref{Avgv2sq}), we can show if $T_{1}>T_{2}$, then $T_{1}>T_{v_{1}}$ ($T_{2}<T_{v_{2}}$). Thus, from eqs. (\ref{heatrate_left_2Bead}) and (\ref{heatrate_right_2Bead}), we get $\dot{Q}_{1}>0$ ($\dot{Q}_{2}<0$). From eqs. (\ref{Avgv1sq}), (\ref{Avgv2sq}) and from law of energy conservation at each bead (Eq. (\ref{Qdot21})), we can write 
\begin{equation}
  \dot{Q}_{1} = -\dot{Q}_{2} =   
 \frac{\kappa^{2}(T_{1}-T_{2})}{\Delta} = \dot{Q}_{21} = -\dot{Q}_{12} > 0,
 \label{Heat_flux_LR_2Bead}
\end{equation}
where $\dot{Q}_{21} (\dot{Q}_{12})$ is the heat flux  from bead $1$ to $2$ ($2$ to $1$), and 
\begin{align}
\Delta =& \; \; (k_{2}+\kappa)\gamma_{1}+(k_{1}+\kappa)\gamma_{2}
+\left(\frac{m_{1}}{\gamma_{1}}+\frac{m_{2}}{\gamma_{2}}\right)\kappa^{2}\nonumber \\
&+\frac{((k_{2}+\kappa)m_{1}-(k_{1}+\kappa)m_{2})^{2}}{m_{2}\gamma_{1}+m_{1}\gamma_{2}} \ge 0.
\label{Delta}
\end{align}
The total entropy produced, over a time interval $\tau$, 
is given by 
\begin{equation}
\Delta S_{\tau}^{tot}=-\tau\left ( \frac{\dot{Q}_{1}}{T_{1}} + \frac{\dot{Q}_{2}}{T_{2}} \right)=\frac{\tau \kappa^{2}(T_{1}-T_{2})^{2}}{T_{1}T_{2}\Delta}.
\label{ent_prod_2Beads}
\end{equation}
\begin{figure}
\includegraphics[scale=0.35, angle=270]{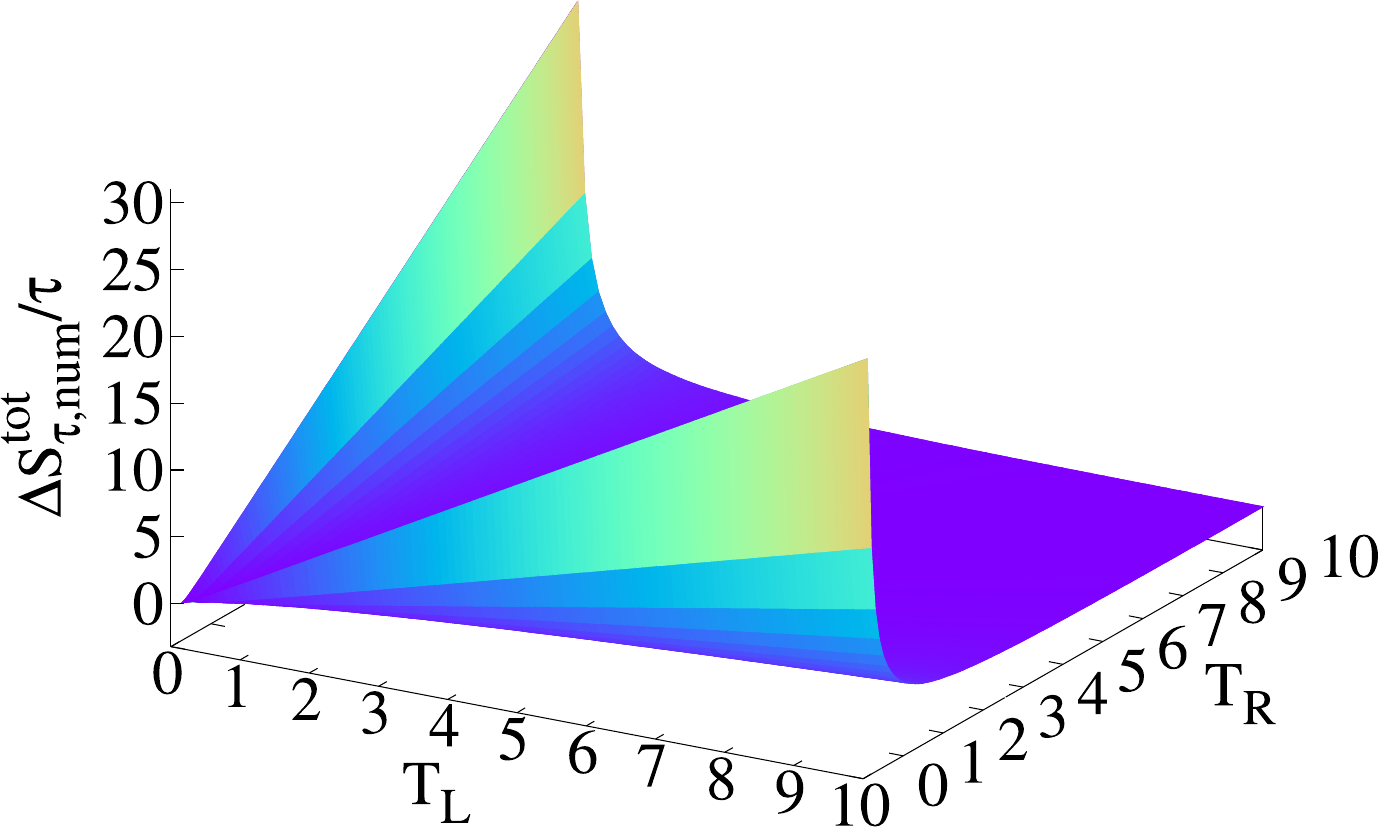}
\caption{The analytically computed rate of entropy production rate (Eq. (\ref{ent_prod_2Beads})) for the temperature of the heat bath towards left ($T_{1}$) and the same towards right ($T_{2}$) for two-beads and the spring model. We have drawn the curves for $\gamma_{1}=\gamma_{2}=1$, $k_{1}=k_{2}=1$, $\kappa = 2$, and $m_{1}=m_{2}=1$.}
\label{Therm_Ent_Prod}
\end{figure}
%
There is a symmetry about $T_{1}=T_{2}$ line (see Fig. \ref{Therm_Ent_Prod}) and the entropy production is zero along the $T_{1}=T_{2}$ line. 
In a symmetrical setting ($k_{1}=k_{2}$ and $\gamma_{1}=\gamma_{2}$), the entropy production 
is invariant under the interchange $m_1 \longleftrightarrow m_2$. 
We get the maximum entropy production when the masses becomes vanishingly small ($m_{1}, m_2\rightarrow 0$), i.e. the overdamped limit. This limit is continuous and does not involve any singular behavior or a coarse graining of the degrees of freedom. Our result matches at this limit with the entropy production in overdamped limit starting from only the position degrees of freedom in the work of Li et. al. \cite{li2019quantifying}. This behavior should be distinguished from approaches in which overdamped dynamics is obtained by coarse graining an underlying underdamped description, for example by integrating out the velocity degrees of freedom, a procedure that can reduce contributions to entropy production from our expression, where the difference is called 'hidden entropy' \cite{celani2012anomalous}. In those coarse-grained descriptions, the static steady-state distributions may converge smoothly to the overdamped limit, while the entropy production does not coincide with the overdamped entropy production \cite{spinney2012entropy,celani2012anomalous}.

On the other extreme, the entropy production vanishes when the coupling constant $\kappa \to 0$, and a bead is in equilibrium with its respective bath. 
Similarly,  if each $\gamma_{\alpha} \to 0$, then $\Delta \rightarrow \infty$, and so $\Delta S_{\tau}^{tot} \rightarrow 0$. 

\subsection{\label{thermal_cond} Thermal conductivity}
We define thermal conductivity ($\lambda$) as the heat flux  between the beads ($\dot{Q}_{21}$) per unit difference of their kinetic temperatures.  
Using eq. (\ref{Heat_flux_LR_2Bead}), we obtain 
\begin{align}
\lambda = \frac{\dot{Q}_{21}}{T_{v_{1}}-T_{v_{2}}} = \frac{\kappa^{2}(T_{1}-T_{2})}{\Delta(T_{v_{1}}-T_{v_{2}})}.
\label{therm_cond}
\end{align}
From eqs. (\ref{Avgv1sq}) and (\ref{Avgv2sq}), we can write 
\begin{align}
T_{v_{1}}-T_{v_{2}} = \frac{(T_{1}-T_{2})}{\Delta}\left(\Delta-\kappa^{2}\left( \frac{m_{1}}{\gamma_{1}} + \frac{m_{2}}{\gamma_{2}}\right)\right).
\label{diff_kin_temp}
\end{align}
Due to $T_1 > T_2$ and eq. (\ref{Delta}), we have 
$T_{v_{1}}- T_{v_{2}} > 0$. From eqs. (\ref{therm_cond}) and (\ref{diff_kin_temp}), we obtain 
\begin{align}
\lambda = \frac{\kappa^{2}}{\left(\Delta-\kappa^{2}\left( \frac{m_{1}}{\gamma_{1}} + \frac{m_{2}}{\gamma_{2}}\right)\right)} \equiv  \frac{\kappa^{2}}{\bar{\Delta}} \geq 0. 
\label{Thermal_cond_lambda}
\end{align}

\begin{figure}
\includegraphics[scale=1.1, angle=270]{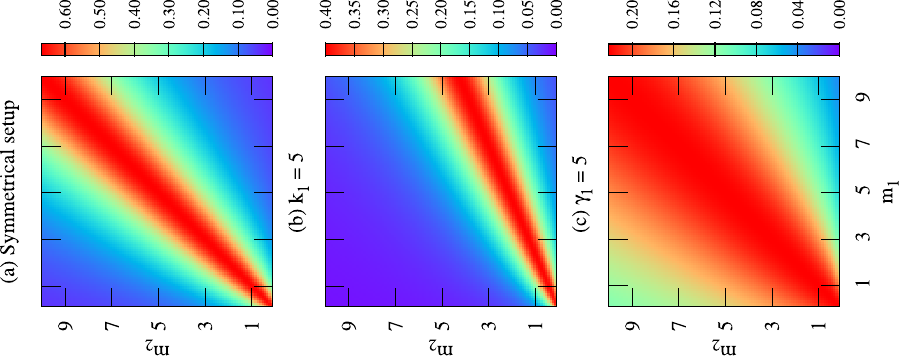}
\caption{Thermal conductivity $\lambda$ [Eq.~(\ref{Thermal_cond_lambda})] in the $(m_1,m_2)$ plane. (a) Symmetric setup: $\gamma_1=\gamma_2=1$, $k_1=k_2=1$, $\kappa=2$. (b) $k_1=5$, $\kappa=2$, all other parameters equal to 1. (c) $\gamma_1=5$, $\kappa=2$, all other parameters equal to 1.}
\label{Therm_Conductivity}
\end{figure}

Note that 
$\lambda$ depends on the system parameters, but 
not on the bath temperatures. For a symmetric setup ($k_1=k_2$, $\gamma_1=\gamma_2$), $\lambda$ is invariant under the exchange $(m_1,m_2)\leftrightarrow(m_2,m_1)$, giving the symmetry about the line $m_1=m_2$, as observed in Fig.~\ref{Therm_Conductivity}(a). Further, the thermal conductivity is maximized for
\begin{eqnarray}
m_{2}=m_1\frac{k_{2}+\kappa}{k_{1}+\kappa}.
\label{m2m1}
\end{eqnarray}
For fixed spring and friction coefficients, eq.~(\ref{m2m1}) determines the optimal mass ratio that maximizes heat transport,
\begin{equation}
\lambda_{\rm max}=\frac{\kappa^{2}}{(k_{2}+\kappa)\gamma_{1}+(k_{1}+\kappa)\gamma_{2}}.
\end{equation}
In the limit $m_{1},m_{2}\rightarrow\infty$, both heat current and entropy production vanish as the dynamics freezes, giving $\lambda\rightarrow0$ for generic mass ratios. A finite conductivity survives only along the special trajectory defined by eq.~(\ref{m2m1}), where the divergent contribution to $\bar{\Delta}$ cancels. By contrast, $k_{1},k_{2}\rightarrow\infty$ suppress transport through confinement, whereas $\gamma_{1},\gamma_{2}\rightarrow\infty$ suppress it through overdamping. Although both limits are insulating, they arise from distinct physical mechanisms.


Figure~\ref{Therm_Conductivity}(a) shows that the conductivity decreases away from the line $m_1=m_2$. When $k_1\neq k_2$, the line of maximum shifts according to eq.~(\ref{m2m1}) [Fig.~\ref{Therm_Conductivity}(b)]. Unequal friction coefficients break the symmetry about the maximum while preserving its location at $m_1=m_2$ [Fig.~\ref{Therm_Conductivity}(c)]. Thus, elastic asymmetry changes the optimal mass-matching condition, whereas dissipative asymmetry primarily modifies the magnitude of transport.

Alternately, we may define conductivity in terms of difference of bath temperatures,
\[
\lambda_{\rm bath}=\frac{\dot Q_{21}}{T_1-T_2}
=\frac{\kappa^2}{\Delta}.
\]
This also obtains a maximum and a symmetry about $m_1=m_2$ line for the symmetric setup. However, unlike eq.~(\ref{Thermal_cond_lambda}), the location of the optimum becomes sensitive to frictional asymmetry. In contrast, $\lambda$ preserves the invariant maximum along $m_1=m_2$, demonstrating that kinetic temperatures act as natural variables governing transport within the present Langevin model. In the infinite-mass limit satisfying eq.~(\ref{m2m1}), $\dot Q_{21}\rightarrow0$ together with $T_{v_1}-T_{v_2}\rightarrow0$, yielding a finite $\lambda_{\rm max}$, whereas $\lambda_{\rm bath}\rightarrow0$. Thus, only the constitutive relation based on kinetic temperatures remains meaningful in this limit. This conclusion is specific to the present model 
may not apply to arbitrary nonequilibrium systems.

For $\kappa=0$, the beads are uncoupled and no heat is transported. Since eq.~(\ref{diff_kin_temp}) gives
$T_1\ge T_{v_1}\ge T_{v_2}\ge T_2$, 
so heat flows from the hotter to the colder reservoir through the kinetic-temperature gradient established between the beads. The required position--velocity correlations vanish when $T_1=T_2$, recovering equilibrium.

\subsection{\label{internal_energy} Equipartition of the mean energy}
The mean internal energy, $\langle E\rangle$, can be expressed as (see eq. (\ref{energy_2beads}))
\begin{align}
\langle E \rangle & = \frac{T_{v_{1}}+T_{v_{2}}}{2} \nonumber \\
 & +\left \langle \frac{1}{2}(k_{1}+\kappa)x_{1}^{2}+\frac{1}{2}(k_{2}+\kappa)x_{2}^{2}-\kappa x_{1}x_{2}\right \rangle\label{energy_equipartition}.
\end{align}
On the other hand, using the covariance elements relationships via eqs. (\ref{Avgx1sq}), (\ref{Avgx2sq}), (\ref{Avgx1x2}),  (\ref{Avgv1sq}), and (\ref{Avgv2sq}), 
we can show that
\begin{align}
\left \langle \frac{1}{2}(k_{1}+\kappa)x_{1}^{2}+\frac{1}{2}(k_{2}+\kappa)x_{2}^{2}-\kappa x_{1}x_{2}\right \rangle &= \frac{T_{v_{1}}+T_{v_{2}}}{2}\label{modetemp_rel_kintemp}.
\end{align}
Combining the above two equations, we have
\begin{align}
\langle E \rangle &= T_{v_{1}}+T_{v_{2}}.
\label{energy_kintemp}
\end{align}
 In this sense, we have a generalized equipartition relation in which the mean energy of the two-beads system is expressed in terms of the kinetic temperatures.

The mean energy is also expressed as 
$\langle E \rangle = 2 T_e$ (see eq.  (\ref{int_eng_wtemp})). This implies 
$T_{e}=({T_{v_{1}}+T_{v_{2}}})/{2}$.
 Equivalently, we can write 
$T_e=C_1T_1+(1-C_{1})T_{2}$, 
where the weight $C_1$ is given by (see Appendix A)
\begin{eqnarray}
 C_{1} = \frac{2\kappa^{2}m_{2}\gamma_{1}(m_{2}\gamma_{1}+m_{1}\gamma_{2})+\gamma_{1}\gamma_{2}\Omega}{2\kappa^{2}(m_{2}\gamma_{1}+m_{1}\gamma_{2})^{2}+2\gamma_{1}\gamma_{2}\Omega}, 
\label{Coeff_Calc}
\end{eqnarray}
with $\Omega = [m_{2}(k_{1}+\kappa)-m_{1}(k_{2}+\kappa)]^{2}+(m_{2}\gamma_{1}+m_{1}\gamma_{2})[(k_{1}+\kappa)\gamma_{2}+(k_{2}+\kappa)\gamma_{1}]$.
Thus, our result generalizes the expression 
for $T_e$ as found in Ref.~\cite{tu2025weighted}, 
which was restricted to the case of equal masses.

It is interesting that in the strong-coupling limit ($\kappa\rightarrow\infty$),  $C_1$ depends only on the ratio of the bead masses, and given by
\begin{eqnarray}
C_{1} = \frac{2m_{2}\gamma_{1}(m_{2}\gamma_{1}+m_{1}\gamma_{2})+\gamma_{1}\gamma_{2}(m_{2}-m_{1})^{2}}{2(m_{2}\gamma_{1}+m_{1}\gamma_{2})^{2}+2\gamma_{1}\gamma_{2}(m_{2}-m_{1})^{2}}.
\label{Te_kinfty}
\end{eqnarray}
In this limit, with equal masses, we obtain 
\begin{equation}
T_e=\frac{\gamma_1T_1+\gamma_2T_2}{\gamma_1+\gamma_2},
\end{equation}
which is actually the effective temperature for a single bead in contact with the two baths \cite{tu2025weighted}. 
Similarly,  the limiting cases with asymmetric masses 
yield
\begin{equation}
T_{e} = \frac{\gamma_{1}T_{1}+(2\gamma_{2}+\gamma_{1})T_{2}}{2(\gamma_{1}+\gamma_{2})} \quad \text{for} \quad m_{1}\gg m_{2}
\label{Te_m1ggm2}
\end{equation}
and
\begin{equation}
T_{e} = \frac{(2\gamma_{1}+\gamma_{2})T_{1}+\gamma_{2}T_{2}}{2(\gamma_{1}+\gamma_{2})} \quad \text{for} \quad m_{2}\gg m_{1}.
\label{Te_m2ggm1}
\end{equation}
This leads us to the 
observation that unlike transport observables, the coefficient $C_1$ is path dependent in both the overdamped and infinite-mass limits, since there  the overall mass scale is absent and only the ratio $m_1/m_2$ matters. Consequently, the effective temperature and mean energy retain memory of the relative mass distribution. In contrast, the entropy production, the heat current, and the thermal conductivity possess unique limiting values,  independent of the approach to the limit. Thus, we may state that the energy-storage observables exhibit a path dependence, whereas measures 
of transport and irreversibility do not.

Equation (\ref{m2m1}) implies that the maximum conductivity is obtained for nearly equal masses in the strong-coupling limit. Here $\lambda_{\rm max}\sim\kappa/(\gamma_1+\gamma_2)$ diverges although both the heat current and entropy production remain finite. The divergence arises because the kinetic-temperature difference vanishes while the heat current remains finite, and is therefore analogous to divergent response coefficients in ballistic transport and superconductivity \cite{ashcroft1976solid,rieder1967properties,dhar2008heat,tinkham2004introduction}. A second divergence occurs when $\gamma_\alpha\rightarrow0$, where the system decouples from the reservoirs. In this case both heat current and entropy production vanish, while the conductivity diverges because the kinetic-temperature difference approaches zero even faster. Thus the two divergences in $\lambda$ originate from distinct physical mechanisms: strong coupling
in the former case, and vanishing dissipation in the latter. \\

In general, $C_1\neq1/2$. Only for equal masses and equal friction coefficients do
we obtain $C_1 = 1/2$, or 
$T_e=(T_1+T_2)/{2}$.
 Remarkably, Table~\ref{tab_limit_beh} shows that the mean energy and effective temperature remain finite in all limiting regimes, even when transport coefficients vanish or diverge. This reflects the fundamentally different nature of energy storage compared with transport and irreversibility.

\begin{table}
\centering
\setlength{\tabcolsep}{3pt}
\begin{tabular}{rccccc}
\hline
\hline
Limit taken & $\langle E \rangle$ & $\dot{Q}_{21}$ & $\Delta S^{tot}_{\tau}$ & $\lambda_{max}$ & $T_{v_{1}}^{max}-T_{v_{2}}^{max}$\\
\hline
$\kappa \rightarrow \infty$ & finite & finite & finite & $\infty$ & 0\\
$m_{1},m_{2} \rightarrow 0$ & finite & finite & finite & finite & finite\\
$m_{1},m_{2} \rightarrow \infty$ & finite & 0 & 0 & finite & 0\\
$\gamma_{1},\gamma_{2} \rightarrow \infty$ & finite & 0 & 0 & 0 & finite\\
$k_{1}, k_{2} \rightarrow 0$ & finite & finite & finite & finite & finite\\
$k_{1},k_{2} \rightarrow \infty$ & finite & 0 & 0 & 0 & finite\\
\hline
\hline
\end{tabular}
\caption{Limiting behavior of energy, heat flux, entropy production, maximum thermal conductivity, and kinetic-temperature difference in the two-bead model. The masses satisfy eq.~(\ref{m2m1}) whenever $\rm max$ is quoted.}
\label{tab_limit_beh}
\end{table}
Unlike entropy production and thermal conductivity, the mean energy and effective temperature depend on the arrangement of unequal masses even in a symmetric setup. Since
interchanging the bead masses changes the weighting coefficient $C_1$ while leaving the bath temperatures fixed. Consequently, the redistribution of inertia modifies the stored energy without affecting the transport observables.


\subsection{\label{ergodicity} Ergodicity}
\begin{figure}
\includegraphics[scale=0.3, angle=270]{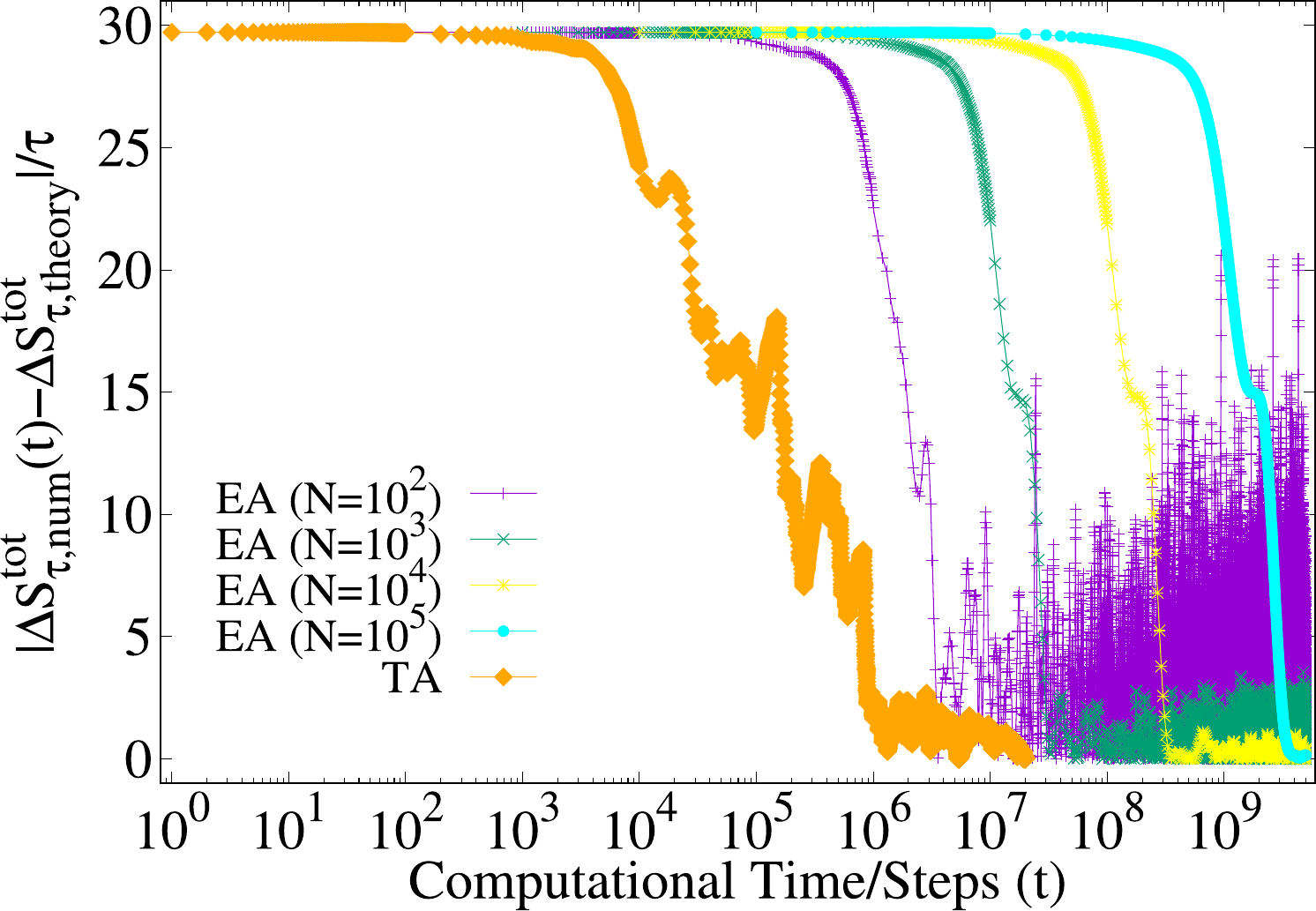}
\caption{Absolute deviation of the entropy production rate from the theoretical value for ensemble averaging (EA) with different numbers of trajectories and time averaging (TA). TA converges with fewer computational steps, whereas EA requires longer simulations but exhibits progressively smaller fluctuations as $N$ increases. Parameters: $T_1=99$, $T_2=1$, $\gamma_1=\gamma_2=1$, $k_1=k_2=1$, $\kappa=2$, $m_1=m_2=1$.}
\label{EnsvsTS_Ent_Prod}
\end{figure}

Our system is governed by linear Langevin dynamics with additive Gaussian noise satisfying the fluctuation--dissipation relation, for which ergodicity is expected and can be established rigorously under standard conditions \cite{frank2005nonlinear,eckmann1999non,risken1996Fokker,maes2003time}.
To confirm this expectation, we compared ensemble averaging (EA) over $N$ independent Langevin trajectories with time averaging (TA) along a single trajectory. The Langevin equations are integrated using the Euler--Maruyama method \cite{nayak2021numerical,erfanian2016using,bayram2018numerical} with $dt=10^{-4}$ from the initial condition $(x_i,v_i)=(0,0)$. Figure~\ref{EnsvsTS_Ent_Prod} compares the convergence of the entropy production obtained from the two averaging protocols.
The orange curve in Fig. \ref{EnsvsTS_Ent_Prod} represents the case for TA protocol reaching the steady state at the longest time ($t/N$), but with the least fluctuations as compared to the other curves from EA even though they appear to have reached steady state earlier, they have significantly larger fluctuations due to finite samples which reduces with more number of parallel trajectories. Therefore, the probability distributions studied in this paper are derived from time series having at least $10^{7}$ samples and the average is taken over this time series as this would need a smaller computational time.

The difference in the apparent convergence times does not reflect different relaxation dynamics, but different statistical convergence properties. Ensemble averages follow the instantaneous relaxation of the mean observable and therefore approach the steady state exponentially on the relaxation timescale $\tau$. In contrast, time averages involve the entire trajectory, including early-time transients, resulting in an algebraic convergence. After relaxation, the residual fluctuations in EA arise from finite-sample statistics and scale with the number of trajectories, whereas TA suppresses fluctuations by increasing the trajectory length. This distinction is statistical rather than physical and applies equally to equilibrium and nonequilibrium systems with finite relaxation times.

\begin{figure}
\includegraphics[scale=0.35, angle=270]{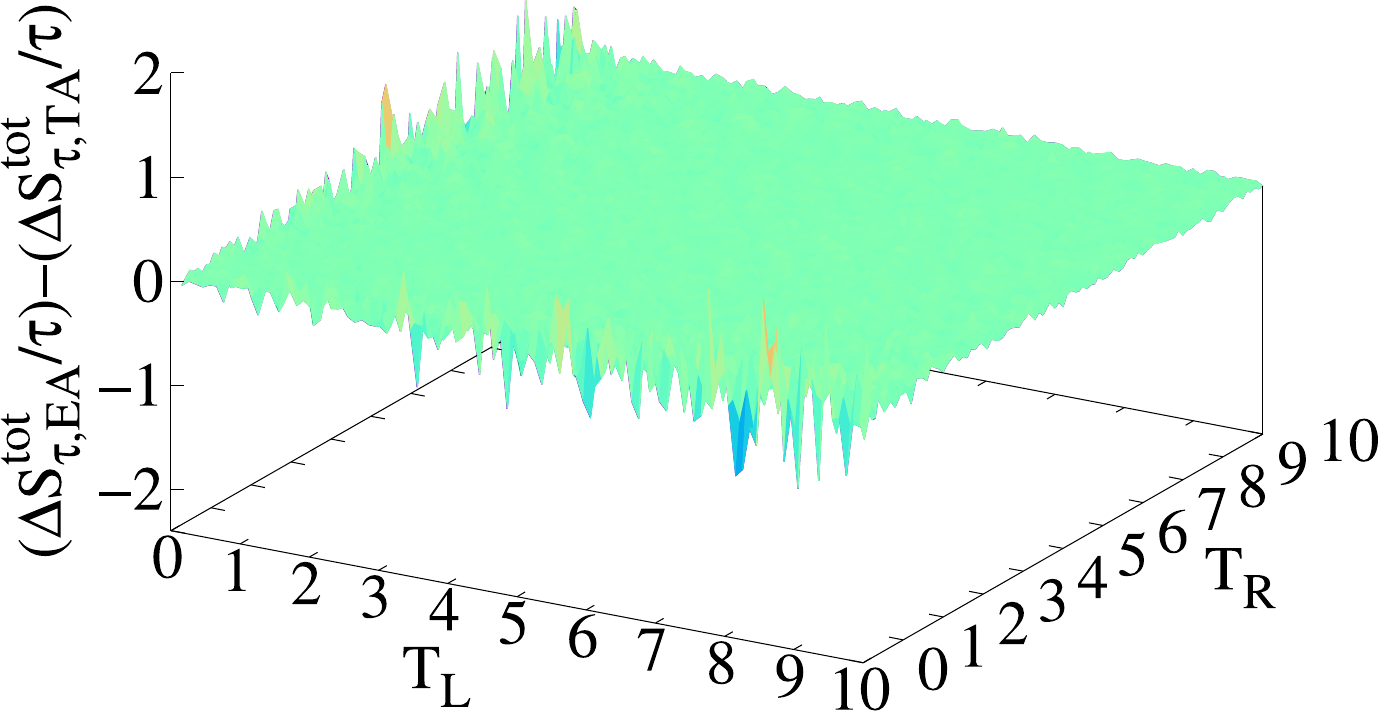}
\caption{Comparison of the entropy production rate obtained from ensemble averaging and time averaging. The negligible difference demonstrates ergodicity. Parameters are the same as in Fig.~\ref{EnsvsTS_Ent_Prod}.}
\label{Dev_Ent_Prod_EAandTA}
\end{figure}
%
%
%

\begin{figure}
\centering{\includegraphics[scale=0.35]{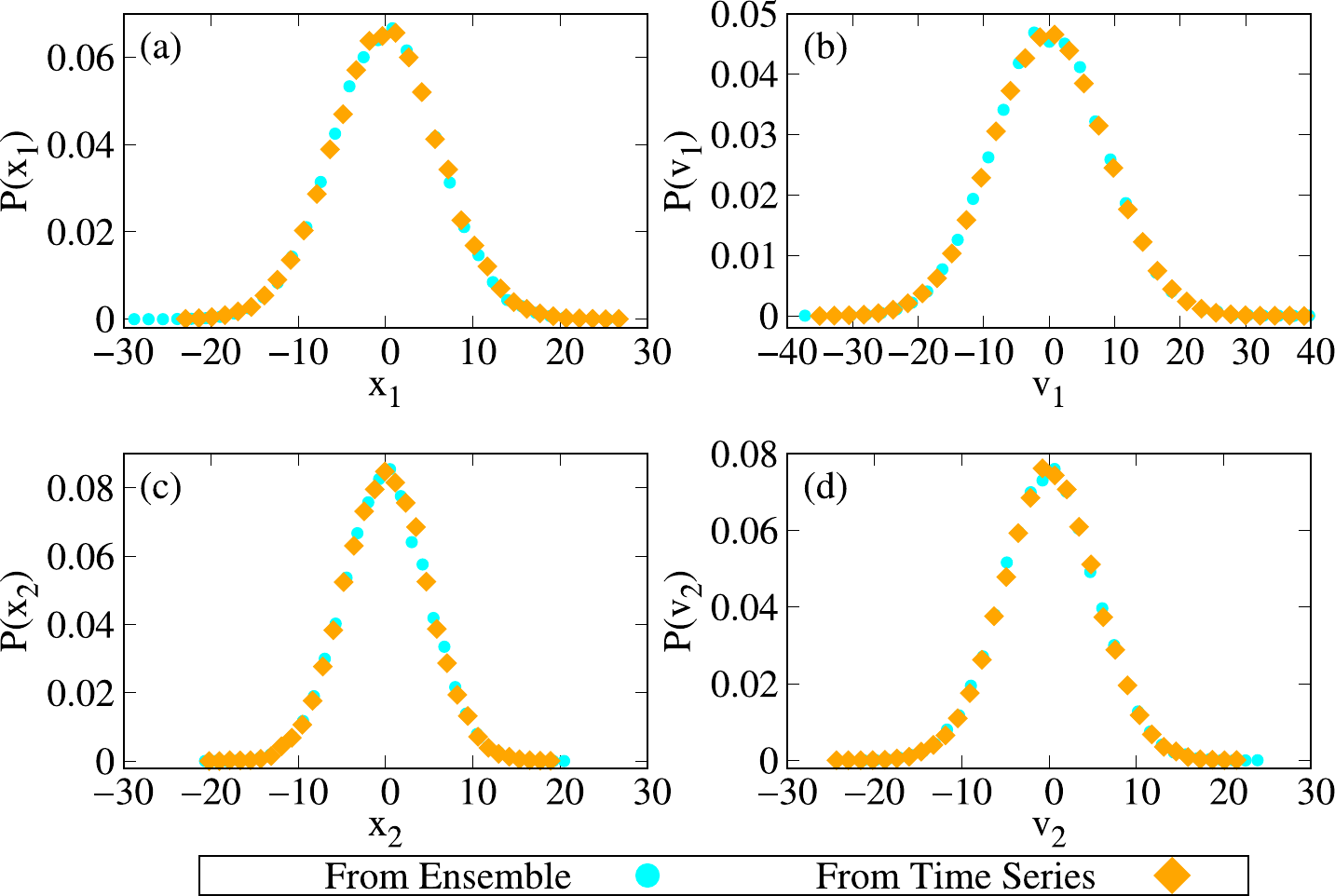}}
\caption{Marginal position and velocity distributions obtained from ensemble averaging and time averaging for the two-bead system. The two curves overlap for all degrees of freedom, confirming ergodicity. Parameters are the same as in Fig.~\ref{EnsvsTS_Ent_Prod}.}
\label{Prob_TwoBead_EAandTA}
\end{figure}

Figure~\ref{Dev_Ent_Prod_EAandTA} compares the entropy production rate obtained from ensemble and time averaging. The difference between the two estimates is negligible, demonstrating the equivalence of the two averaging procedures for this observable. We further compare the marginal distributions of all positions and velocities obtained from ensemble and time averaging. As shown in Fig.~\ref{Prob_TwoBead_EAandTA}, the distributions are indistinguishable, providing additional evidence of ergodicity. The agreement between ensemble and time averages for both the entropy production rate (Fig.~\ref{Dev_Ent_Prod_EAandTA}) and all marginal distributions (Fig.~\ref{Prob_TwoBead_EAandTA}) provides a strong numerical evidence for ergodicity.

\section{\label{Discussions} Conclusions}

We combined numerical simulations of Langevin dynamics with covariance-matrix theory to analyze a two-bead nonequilibrium system. The steady-state distributions of positions doesn't depend explicitly on mass arrangement in a symmetrical setup and reduce smoothly to overdamped results in the zero-mass limit \cite{li2019quantifying}.

Entropy production is maximized in the overdamped regime and vanishes in the absence of coupling or dissipation. The zero-mass limit reproduces overdamped behavior but remains distinct from coarse-graining due to hidden entropy contributions \cite{celani2012anomalous}. The approach to steady state and relaxation pathways may be further analyzed along the lines of Ref.~\cite{netz2020approach}.

Kinetic temperatures emerge as the natural variables governing the system. Their difference controls heat transport and entropy production, while their sum determines the mean energy through a generalized equipartition relation. 

Thermal conductivity defined in terms of kinetic temperatures provides a consistent constitutive description of heat transport. In this formulation, the heat current is proportional to the difference of bead kinetic temperatures, in analogy with Fourier’s law. The resulting conductivity exhibits a symmetric maximum along $m_1=m_2$ in the $(m_1,m_2)$ plane, reflecting balanced inertial effects. This symmetry is broken when asymmetries in friction coefficients or coupling strengths are introduced, highlighting the sensitivity of transport coefficients to dynamical parameters. In contrast, definitions based solely on bath temperatures do not capture this structure, emphasizing that kinetic temperatures encode the relevant internal degrees of freedom governing transport.

In the infinite-mass limit, bath-temperature differences fail to define a consistent driving force, whereas kinetic-temperature differences vanish together with the current as shown in Table \ref{tab_limit_beh}, yielding a well-defined constitutive relation. This identifies kinetic temperatures as the appropriate variables governing transport. Divergent conductivities arise from distinct physical mechanisms---either strong coupling or vanishing dissipation---highlighting that large response does not necessarily imply enhanced transport.

Across limiting regimes, entropy production and heat current behave similarly, while mean energy remains finite, also shown in Table \ref{tab_limit_beh}, revealing a separation between irreversibility and energy storage. Mass exchange alters energy partitioning but leaves transport invariant in a symmetrical setup, demonstrating that inertia selectively affects time-symmetric observables without modifying irreversibility. The effective temperature 
has been explicitly derived for the case of unequal masses, which extends the results of Ref.~\cite{tu2025weighted} with equal masses. The present results rely on linear interactions and Gaussian steady states. 
For nonlinear interactions, steady-state distributions are generally non-Gaussian, and relations such as generalized equipartition and mass-exchange invariance are not expected to hold in the same form. 
In Hamiltonian systems without stochastic baths, kinetic temperatures do not uniquely characterize nonequilibrium states, and transport is governed by different mechanisms \cite{dhar2008heat, lepri2003thermal}.

These predictions can be tested in mesoscopic nonequilibrium systems such as optically trapped colloidal particles and micromechanical or nanomechanical resonators \cite{Blickle2012,Martinez2016,Vinante2008,Gieseler2013}. In such platforms, trap stiffness, damping, and coupling can be tuned independently, while effective masses can be varied via particle composition or attached microstructures. Measurements of stochastic heat currents and velocity fluctuations would allow direct verification of (i) the invariance of entropy production under mass exchange, (ii) the path dependence of energy observables, and (iii) the role of kinetic temperatures as the relevant transport variables, particularly in extreme limits such as large mass or strong coupling.

Future directions include extensions to multi-bead harmonic networks, where kinetic temperatures may lead to mode-dependent generalized equipartition and transport optimization manifolds \cite{dhar2008heat,lebowitz1999gallavotti}. It would be important to test the robustness of the separation between energy storage and irreversibility in anharmonic or non-Gaussian systems \cite{jarzynski2011annu}. The central role of kinetic temperatures also suggests applications to active and driven systems, where bath temperatures are ill-defined \cite{seifert2012stochastic}. Finally, our results motivate a deeper investigation of hidden entropy production and its information-theoretic interpretation under coarse-graining \cite{spinney2012entropy,esposito2012stochastic}, as well as systematic experimental validation in controllable mesoscopic platforms \cite{bechinger2016active}. Owing to its simplicity as a minimal model for heterogeneous nonequilibrium environments, the two-bead system provides a foundation for systematically exploring the thermodynamics of more complex heterogeneous networks and landscapes.

\appendix

\renewcommand{\theequation}{\Alph{section}.\arabic{equation}}

\makeatletter
\@addtoreset{equation}{section}
\makeatother

\section{\label{Appendix A} Weighted Effective Temperature}
The Langevin equations 
(\ref{pos1_Langevin})-(\ref{vel2_Langevin}) could be
compactly  written as
\begin{equation}
\frac{d\boldsymbol{z}}{dt} = - \boldsymbol{Az} + \boldsymbol{\xi}, \label{vec_Langevin}
\end{equation}
where for the two-beads system  $\boldsymbol{z} = [x_{1},x_{2},v_{1},v_{2}]^{T}$ and $\boldsymbol{\xi} = [0, 0, \xi_{1}, \xi_{2}]^{T}$. The coefficient matrix is 
given by
\begin{equation}
\boldsymbol{A} =
\begin{bmatrix}
0 & 0 & -1 & 0 \\
0 & 0 & 0 & -1 \\
k_{1}/m_{1} & -\kappa/m_{1} & \gamma_{1}/m_{1} & 0\\
-\kappa/m_{2} & k_{2}/m_{2} & 0 & \gamma_{2}/m_{2}\\
\end{bmatrix}.
\label{coeff_matrix}
\end{equation}
 The effective spring constants $k_{1}+\kappa$ will be denoted as $k_{1}$ for the leftmost particle and $k_{2}+\kappa$ as $k_{2}$ for the rightmost particle for the discussion in Appendix A and B.

The covariance matrix ($\boldsymbol{\sigma}$) will be a symmetric $4 \times 4$ matrix. And the modified matrices $\tilde{\boldsymbol{\sigma}}$ 
and $\tilde{\boldsymbol{A}}$ will satisfy 
\begin{equation}
\tilde{\boldsymbol{\sigma}} = \boldsymbol{\sigma} \tilde{\boldsymbol{K}} = \boldsymbol{\sigma}\begin{bmatrix}
\boldsymbol{K} & \boldsymbol{0}\\
\boldsymbol{0} & \boldsymbol{M}
\end{bmatrix} =  \begin{bmatrix}
\boldsymbol{\sigma_{xx}}\boldsymbol{K} & \boldsymbol{\sigma_{xv}M}\\
\boldsymbol{\sigma_{vx}}\boldsymbol{K} & \boldsymbol{\sigma_{vv}M}
\end{bmatrix},
\label{Sigmatilde}
\end{equation}
where $\boldsymbol{v} = [v_{1},v_{2}]^{T}$ and $\boldsymbol{x} = [x_{1},x_{2}]^{T}$. 
Similarly, 
\begin{equation}
\tilde{\boldsymbol{A}} = 
\tilde{\boldsymbol{K}}\boldsymbol{A} = 
\begin{bmatrix}
\boldsymbol{0} & -\boldsymbol{K}\\
\boldsymbol{K} & \boldsymbol{\Gamma}
\end{bmatrix},
\label{Atilde}
\end{equation}
where $\boldsymbol{K} =  \begin{bmatrix}
k_{1} & -\kappa\\
-\kappa & k_{2}
\end{bmatrix}$, $\boldsymbol{M} =  \begin{bmatrix}
m_{1} & 0\\
0 & m_{2}
\end{bmatrix}$ and $\boldsymbol{\Gamma} =  \begin{bmatrix}
\gamma_{1} & 0\\
0 & \gamma_{2}
\end{bmatrix}$.
$\tilde{\boldsymbol{\sigma}}$ is assumed to satisfy the following relation \cite{tu2025weighted}
\begin{equation}
\tilde{\boldsymbol{\sigma}} = T_{e}\boldsymbol{I}_{4} + \tilde{\boldsymbol{\sigma}}^{r},
\label{sigmatilde}
\end{equation}
where $\text{Tr}\tilde{\boldsymbol{\sigma}}^{r}=0$ and $\boldsymbol{I}_{4}$ is the $4\times4$ identity matrix. 
Again, from the Lyapunov equation for the two-beads setup 

\begin{equation}
\boldsymbol{A\sigma}+\boldsymbol{\sigma A}^{T} = 2\boldsymbol{D}.
\label{CT_Lyapunov}
\end{equation}

and the ansatz [Eq. (\ref{sigmatilde})], 
we find that the residual matrix must satisfy
\begin{gather}
\tilde{\boldsymbol{A}}\tilde{\boldsymbol{\sigma}}^{r}+\tilde{\boldsymbol{\sigma}}^{r T} \tilde{\boldsymbol{A}}^{T} =  2\left(\tilde{\boldsymbol{K}}.\boldsymbol{D}.\tilde{\boldsymbol{K}}-T_{e}\begin{bmatrix}
\boldsymbol{0} & \boldsymbol{0} \\
\boldsymbol{0} & \boldsymbol{\Gamma}
\end{bmatrix}\right) \nonumber \\
 = 2\begin{bmatrix}
0 & 0 & 0 & 0\\
0 & 0 & 0 & 0\\
0 & 0 & \gamma_{1}(T_{1}-T_{e}) & 0 \\
0 & 0 & 0 & \gamma_{2}(T_{2}-T_{e})
\end{bmatrix},
\label{CT_Lyapunov_w_Atilde}
\end{gather}
where 
$\boldsymbol{D}=
\begin{bmatrix} 
0 & 0 & 0 & 0 \\
0 & 0 & 0 & 0 \\
0 & 0 & \gamma_{1}T_{1}/m_{1}^{2}& 0\\
0  & 0 & 0 & \gamma_{2}T_{2}/m_{2}^{2} 
\end{bmatrix}$ is the diffusion matrix for 
the two-beads system. The linear equation (\ref{CT_Lyapunov_w_Atilde}) could be expressed as a linear combination of the two bases $\tilde{\boldsymbol{\sigma}}_{\alpha}$ where $\alpha = 1, 2$ which satisfy 
\begin{equation}
\tilde{\boldsymbol{A}}\tilde{\boldsymbol{\sigma}}_{\alpha}+\tilde{\boldsymbol{\sigma}}_{\alpha}^{T} \tilde{\boldsymbol{A}}^{T} = 2\begin{bmatrix}
0 & 0 & 0 & 0\\
0 &  0 & 0 & 0\\
0 & 0 & \delta_{\alpha 1} & 0 \\
0 & 0 & 0 & \delta_{\alpha 2}
\end{bmatrix}.
\label{CT_Lyapunov_w_base}
\end{equation}
To obtain $\tilde{\boldsymbol{\sigma}}_{\alpha}$
from the above equation, we express them
in the following form:
\begin{equation}
\tilde{\boldsymbol{\sigma}}_{\alpha} = \begin{bmatrix}
\boldsymbol{B}_{\alpha} & \boldsymbol{J}_{\alpha} \\
\boldsymbol{F}_{\alpha} &  \boldsymbol{G}_{\alpha}
\end{bmatrix},
\label{matrixform_base}
\end{equation}
where  $\boldsymbol{B}_{\alpha}$, $\boldsymbol{F}_{\alpha}$, $\boldsymbol{J}_{\alpha}$ and $\boldsymbol{G}_{\alpha}$ are $2\times2$ matrices. The symmetry conditions on the modified covariance matrix ($\tilde{\boldsymbol{\sigma}}$) in eq. (\ref{Sigmatilde}) are assumed to be preserved for the basis matrices ($\tilde{\boldsymbol{\sigma}}_{\alpha}$) as well and the solutions are computed based on it. Together with these conditions and eq. (\ref{CT_Lyapunov_w_base}), we get the following matrix relations
\begin{gather}
\boldsymbol{G}_{\alpha}^{T} = \boldsymbol{M}\boldsymbol{G}_{\alpha}\boldsymbol{M}^{-1},  \label{G_matrix} \\
\boldsymbol{F}_{\alpha}^{T} = \boldsymbol{KJ}_{\alpha}\boldsymbol{M}^{-1},\\
(\boldsymbol{KB}_{\alpha})^{T} = \boldsymbol{KB}_{\alpha},\\
(\boldsymbol{KF}_{\alpha})^{T} = -\boldsymbol{KF}_{\alpha},\\ 
\boldsymbol{KB}_{\alpha}+\boldsymbol{\Gamma F}_{\alpha} = \boldsymbol{G}_{\alpha}^{T}\boldsymbol{K},   \\
\boldsymbol{MF}_{\alpha}+\boldsymbol{F}_{\alpha}^{T}\boldsymbol{M}+\boldsymbol{\Gamma G}_{\alpha}+\boldsymbol{G}_{\alpha}^{T}\boldsymbol{\Gamma} = 2\boldsymbol{E}_{\alpha}, \label{E_matrix}
\end{gather}
where $\boldsymbol{E}_{\alpha} =  \begin{bmatrix}
\delta_{\alpha 1} & 0\\
0 & \delta_{\alpha 2}
\end{bmatrix}$. The residual matrix may be expressed as
\begin{equation}
\tilde{\boldsymbol{\sigma}}^{r} = \gamma_{1}(T_{1}-T_{e})\tilde{\boldsymbol{\sigma}}_{1} + \gamma_{2}(T_{2}-T_{e})\tilde{\boldsymbol{\sigma}}_{2}.
\label{Sigma_r}
\end{equation}
By taking the trace of $\tilde{\boldsymbol{\sigma}}^{r}$ in the above eq. (\ref{Sigma_r}) and using the traceless property, we can obtain the effective temperature 
as $T_{e} = C_{1}T_{1}+(1-C_{1})T_{2}$, where 
\begin{equation}
C_{\alpha} = \frac{\gamma_{\alpha}\text{Tr}\tilde{\boldsymbol{\sigma}}_{\alpha}}{\gamma_{1}\text{Tr}\tilde{\boldsymbol{\sigma}}_{1}+\gamma_{2}\text{Tr}\tilde{\boldsymbol{\sigma}}_{2}}.
\label{Coeff_Sigmar}
\end{equation}
The steady state distribution is 
\begin{equation}
P(\boldsymbol{z}) = \exp^{-(H+\Delta H)/T_{e}}/\mathcal{Z},   
\label{Pz}
\end{equation}
where the Hamiltonian ($H$)  is
\begin{eqnarray}
H = \frac{1}{2}\boldsymbol{v}^{T}\boldsymbol{Mv}+\frac{1}{2}\boldsymbol{x}^{T}\boldsymbol{K}\boldsymbol{x} = \frac{1}{2}\boldsymbol{z}^{T}\tilde{\boldsymbol{K}}\boldsymbol{z},
\label{Hamiltonian_2B}
\end{eqnarray}
and the additional term ($\Delta H$) is
\begin{eqnarray}
\Delta H = \frac{1}{2}\boldsymbol{z}^{T}\tilde{\boldsymbol{K}}\left[(\boldsymbol{I}_{4}+\tilde{\boldsymbol{\sigma}}^{r}/T_{e})^{-1}-\boldsymbol{I}_{4} \right]\boldsymbol{z}.
\label{dH_2B}
\end{eqnarray}
We can solve for $\tilde{\boldsymbol{\sigma}}_{\alpha}$ using eqs. (\ref{G_matrix}-\ref{E_matrix}). Then, we get $\tilde{\boldsymbol{\sigma}}^{r}$ from eq. (\ref{Sigma_r}) .  Thus, if we know $\tilde{\boldsymbol{\sigma}}_{\alpha}$, we can determine $\tilde{\boldsymbol{\sigma}}^{r}$, $C_{\alpha}$ and $T_{e}$. The coefficients come out as given in (\ref{Coeff_Calc}). We get a non-zero $\Delta H \neq 0$ in $P(z)$ making it non-Boltzmann type, when the system is in NESS ($T_{1}\neq T_{2}$).
\vspace{0.2 cm}


\section{\label{Appendix B} Covariance Matrix Elements}
The elements of the covariance matrix ($\boldsymbol{\sigma}$) could be obtained by solving the linear equations from the Lyapunov equation in eq. (\ref{CT_Lyapunov}). These relations will be useful in deriving useful physical quantities in the main paper. In the following, $\langle...\rangle$ stands for
an average over $P(z)$ (Eq. (\ref{Pz})).

\begin{widetext}
\begin{gather}
\sigma_{x_{1}x_{1}} = \langle x_{1}^{2} \rangle = \frac{k_{2}T_{1}}{k_{1}k_{2}-\kappa^{2}}-\frac{\kappa^{2}(T_{1}-T_{2})\left(\frac{k_{2}m_{1}}{\gamma_{1}}+\gamma_{2}+\frac{m_{2}(k_{1}m_{2}-k_{2}m_{1})}{m_{2}\gamma_{1}+m_{1}\gamma_{2}}\right)}{(k_{1}k_{2}-\kappa^{2})\Delta}, \label{Avgx1sq}\\
\sigma_{x_{2}x_{2}} =\langle x_{2}^{2} \rangle = \frac{k_{1}T_{2}}{k_{1}k_{2}-\kappa^{2}}+\frac{\kappa^{2}(T_{1}-T_{2})\left(\frac{k_{1}m_{2}}{\gamma_{2}}+\gamma_{1}+\frac{m_{1}(k_{2}m_{1}-k_{1}m_{2})}{m_{2}\gamma_{1}+m_{1}\gamma_{2}}\right)}{(k_{1}k_{2}-\kappa^{2})\Delta}, \label{Avgx2sq}\\
\sigma_{x_{1}x_{2}} =\langle x_{1}x_{2} \rangle = \frac{\kappa T_{1}}{k_{1}k_{2}-\kappa^{2}}-\frac{\kappa(T_{1}-T_{2})\left(k_{1}\gamma_{2}+\frac{m_{1}\kappa^{2}}{\gamma_{1}}+\frac{k_{1}m_{2}(k_{1}m_{2}-k_{2}m_{1})}{m_{2}\gamma_{1}+m_{1}\gamma_{2}}\right)}{(k_{1}k_{2}-\kappa^{2})\Delta} =\langle x_{2}x_{1} \rangle = \sigma_{x_{2}x_{1}}, \label{Avgx1x2}\\
\sigma_{x_{1}v_{1}} = \langle x_{1}v_{1} \rangle = 0 = \langle v_{1}x_{1} \rangle = \sigma_{v_{1}x_{1}}, \label{Avgx1v1}\\
\sigma_{x_{1}v_{2}} = \langle x_{1}v_{2} \rangle = \frac{\kappa(T_{1}-T_{2})}{\Delta} = \langle v_{2}x_{1} \rangle = \sigma_{v_{2}x_{1}}, \label{Avgx1v2}\\
\sigma_{x_{2}v_{2}} = \langle x_{2}v_{2} \rangle = 0 = \langle v_{2}x_{2} \rangle = \sigma_{v_{2}x_{2}}, \label{Avgx2v2}\\
\sigma_{x_{2}v_{1}} = \langle x_{2}v_{1} \rangle = -\frac{\kappa(T_{1}-T_{2})}{\Delta} = \langle v_{1}x_{2} \rangle = \sigma_{v_{1}x_{2}}, \label{Avgx2v1}\\
\sigma_{v_{1}v_{1}} = \langle v_{1}^{2} \rangle = \frac{T_{1}}{m_{1}}-\frac{\kappa^{2}(T_{1}-T_{2})}{\gamma_{1}\Delta}, \label{Avgv1sq}\\
\sigma_{v_{2}v_{2}} = \langle v_{2}^{2} \rangle = \frac{T_{2}}{m_{2}}+\frac{\kappa^{2}(T_{1}-T_{2})}{\gamma_{2}\Delta}, \label{Avgv2sq}\\
\sigma_{v_{1}v_{2}} = \langle v_{1}v_{2} \rangle = \frac{(m_{1}k_{2}-m_{2}k_{1})\kappa(T_{1}-T_{2})}{(m_{2}\gamma_{1}+m_{1}\gamma_{2})\Delta} = \langle v_{2}v_{1} \rangle = \sigma_{v_{2}v_{1}},
\end{gather}
\end{widetext}

where $\Delta$ is from eq. (\ref{Delta}).

\section{\label{Appendix C} Heat Transport}
There is transfer of heat ($\dot{Q}_{21}$) from the bead $1$ to bead $2$ when the system is coupled to two thermal baths kept at different temperatures. This physical quantity could be obtained from the energy conservation at each beads. We can start from the hamiltonian or the energy ($E$) relation for the two beads and spring system in eq. (\ref{Hamiltonian_2B}) as
\begin{align}
E & = \frac{1}{2}k_{1}x_{1}^{2}+\frac{1}{2}m_{1}v_{1}^{2}+\frac{1}{2}k_{2}x_{2}^{2}+\frac{1}{2}m_{2}v_{2}^{2}+\frac{1}{2}\kappa(x_{1}-x_{2})^{2} \nonumber \\
 & = \frac{1}{2}(k_{1}+\kappa)x_{1}^{2}+\frac{1}{2}m_{1}v_{1}^{2}+\frac{1}{2}(k_{2}+\kappa)x_{2}^{2}+\frac{1}{2}m_{2}v_{2}^{2} \nonumber\\
   & \;\;  -\kappa x_{1}x_{2}. \label{energy_2beads}
\end{align}

There is an interesting property to notice in
Tu's effective temperature written from the internal energy $\langle E\rangle$, derived from eq. (\ref{energy_2beads}) as 

\begin{align}
\langle E\rangle = 
\frac{1}{2}\text{Tr}\left(\begin{bmatrix}
    \langle x_{1}^{2}\rangle & \langle x_{1}x_{2}\rangle \\
    \langle x_{2}x_{1}\rangle & \langle x_{2}^{2}\rangle
\end{bmatrix}
\begin{bmatrix}
    k_{1}+\kappa & -\kappa \\
    -\kappa & k_{2}+\kappa
\end{bmatrix}\right)\nonumber\\ +
\frac{1}{2}\text{Tr}\left(\begin{bmatrix}
    \langle v_{1}^{2}\rangle & \langle v_{1}v_{2}\rangle \\
    \langle v_{2}v_{1}\rangle & \langle v_{2}^{2}\rangle
\end{bmatrix}
\begin{bmatrix}
    m_{1} & 0 \\
    0 & m_{2}
\end{bmatrix}\right).
\label{energy_matrix}
\end{align}
Using eqs. (\ref{Sigmatilde}) and (\ref{sigmatilde}), we get eq. (\ref{int_eng_wtemp}) from eq. (\ref{energy_matrix}).

\begin{align}
\langle E \rangle &= \frac{1}{2}\text{Tr}(\boldsymbol{\sigma_{xx}K})+\frac{1}{2}\text{Tr}(\boldsymbol{\sigma_{vv}M}) \\
 &= \frac{1}{2}\text{Tr}\left(\begin{bmatrix}
 \boldsymbol{\sigma_{xx}K} & 0\\
 0 & \boldsymbol{\sigma_{vv}M}
\end{bmatrix}\right)=\frac{1}{2}\text{Tr}(\tilde{\boldsymbol{\sigma}})\\
 &=\frac{1}{2}\text{Tr}(T_{e}\boldsymbol{I}_{4}+\tilde{\boldsymbol{\sigma}}_{r}) = 2T_{e}.\label{int_eng_wtemp}
\end{align}

We used the ansatz $\text{Tr}(\tilde{\boldsymbol{\sigma}}_{r})=0$ to simplify and get eq. (\ref{int_eng_wtemp}). We have used this relation (Eq. (\ref{int_eng_wtemp})) to establish connection between effective temperature and kinetic temperatures using eq. (\ref{energy_kintemp}).

We can take derivative of energy in eq. (\ref{energy_2beads}) to get

\begin{align} 
\dot{E} & = (k_{1}+\kappa)x_{1}v_{1}+m_{1}\dot{v}_{1}v_{1}+(k_{2}+\kappa)x_{2}v_{2}+m_{2}\dot{v}_{2}v_{2} \nonumber\\
    &\;\;-\kappa(x_{1}v_{2}+x_{2}v_{1}).\nonumber
\end{align}

On taking average,

\begin{align}
\langle \dot{E} \rangle & = (k_{1}+\kappa)\langle x_{1}v_{1}\rangle+\langle m_{1}\dot{v}_{1}v_{1}\rangle+(k_{2}+\kappa)\langle x_{2}v_{2}\rangle \nonumber\\
    & +\langle m_{2}\dot{v}_{2}v_{2}\rangle-\kappa(\langle x_{1}v_{2}\rangle+\langle x_{2}v_{1}\rangle), \nonumber\\
\langle \dot{E} \rangle & = (k_{1}+\kappa)\sigma_{x_{1}v_{1}}+\langle F_{1}v_{1}\rangle+(k_{2}+\kappa)\sigma_{x_{2}v_{2}} \nonumber\\
    & +\langle F_{2}v_{2}\rangle-\kappa(\langle x_{1}v_{2}\rangle+\langle x_{2}v_{1}\rangle),\nonumber
\end{align}

where $F_{1}=m_{1}\dot{v}_{1}$ and $F_{2}=m_{2}\dot{v}_{2}$ is the force on the first and second bead respectively. From the covariance matrix in eqs. (\ref{Avgx1v1}) and (\ref{Avgx2v2}), we get $\sigma_{x_{1}v_{1}}=0$ and $\sigma_{x_{2}v_{2}}=0$.

\begin{align}
\langle \dot{E} \rangle & = \langle F_{1}v_{1}\rangle + \langle F_{2}v_{2}\rangle-\kappa(\langle x_{1}v_{2}\rangle+\langle x_{2}v_{1}\rangle) \nonumber \\
& = \langle P_{1}\rangle + \langle P_{2}\rangle + \langle \dot{U}_{12}\rangle.
\end{align}
Thus, the total rate of energy ($\langle \dot{E} \rangle$) injected into the system could be decomposed into power exerted on the first bead ($\langle P_{1} \rangle = \langle F_{1}v_{1}\rangle$), power exerted on the second bead ($\langle P_{2} \rangle=\langle F_{2}v_{2}\rangle$) and rate of change in energy of the middle spring that is coupling the beads $1$ and $2$ ($\langle \dot{U}_{12} \rangle =-\kappa\langle x_{1}v_{2}+x_{2}v_{1} \rangle$). From the Langevin eqns (\ref{vel1_Langevin}) and (\ref{vel2_Langevin}), one can write $\langle P_{1} \rangle $ and $\langle P_{2} \rangle$.

\begin{align}
\langle P_{1} \rangle & = -(k_{1}+\kappa)\langle x_{1}v_{1}\rangle + \kappa \langle x_{2}v_{1} \rangle -\gamma_{1}\langle v_{1}^{2} \rangle + \langle \xi_{1} v_{1}\rangle, \nonumber\\
\langle P_{2} \rangle& = -(k_{2}+\kappa)\langle x_{2}v_{2}\rangle + \kappa \langle x_{1}v_{2} \rangle -\gamma_{2}\langle v_{2}^{2} \rangle + \langle \xi_{2} v_{2}\rangle. \nonumber
\end{align}

From eqs (\ref{Avgx1v1}), (\ref{Avgx2v2}), (\ref{heatrate_left_2Bead}) and (\ref{heatrate_right_2Bead}), and from these relations $\langle \xi_{1} v_{1}\rangle = \gamma_{1}T_{1}/m_{1}$ and $\langle \xi_{2} v_{2}\rangle = \gamma_{2}T_{2}/v_{2}$ using Novikov’s theorem \cite{novikov1965functionals,van1992stochastic,gardiner2009handbook}. We can get

\begin{align}
\langle P_{1} \rangle & = \kappa \langle x_{2}v_{1} \rangle + \gamma_{1}\left(T_{1}/m_{1} - \langle v_{1}^{2} \rangle \right) = \dot{Q}_{12}+\dot{Q}_{1},\nonumber\\
\langle P_{2} \rangle & = \kappa \langle x_{1}v_{2} \rangle +\gamma_{2}\left(T_{2}/m_{2} - \langle v_{2}^{2} \rangle \right) = \dot{Q}_{21}+\dot{Q}_{2},\nonumber
\end{align}

where the heat ($\dot{Q}_{12}$) transferred from bead $2$ to $1$ is given as $\kappa \langle x_{2}v_{1} \rangle$ and the heat $\dot{Q}_{21}=\kappa\langle x_{1}v_{2}\rangle$ is flowing in the opposite way, they are the excess heat coming besides heat received from the heat baths. Therefore, the power exerted on each bead is the heat transferred from the neighboring heat bath and the heat transferred from the other bead. One can show from the relations of covariance matrix in eqs (\ref{Avgx1v2}), (\ref{Avgx2v1}) and (\ref{Heat_flux_LR_2Bead}), we get $\dot{Q}_{12}=-\dot{Q}_{1}, \dot{Q}_{21}=-\dot{Q}_{2}$ and $\dot{Q}_{12}=-\dot{Q}_{21}$.
\begin{gather}
\dot{Q}_{21}=-\dot{Q}_{12}=\frac{\kappa^{2}(T_{1}-T_{2})}{\Delta},\label{Qdot21}\\
\langle P_{1}\rangle  = 0, \nonumber\\
\langle P_{2}\rangle  = 0, \nonumber\\
\langle \dot{U}_{12} \rangle = 0, \nonumber\\
\langle \dot{E}\rangle = 0.
\end{gather}
Thus, the sign of heat $\dot{Q}_{21}$ ($\dot{Q}_{12}$) being positive (negative) implies that the heat is getting transferred from bead $1$ to $2$ i.e towards the lower temperature from the higher temperature. Moreover, the average power injected at each bead is zero, also implying no net accumulation of heat on each bead at steady state on average. The total average rate of change of energy on the two-beads system is also zero at steady state. 

\bibliography{Paper_BeadSpring_NEq.bib}

\end{document}